\documentclass[11pt]{article}
\usepackage[dvips]{graphicx}  %  XXX archive  version

\begin{document}

\title{{\bf Novel Gravity Probe B Gravitational Wave Detection}}

\author{{ \bf Reginald T. Cahill}\\
School of Chemistry, Physics and Earth Sciences\\
 Flinders University \\
 GPO Box 2100, Adelaide 5001, Australia \\
 Reg.Cahill@flinders.edu.au 
\\ \\ August 21, 2004}

\setcounter{page}{1}

\date{}

\maketitle

%\begin{center} arXiv:physics/0408xx August 2004 \end{center}

\begin{center}{\bf Abstract}   \end{center}
The Gravity Probe B (GP-B) satellite experiment will measure the precession of on-board gyroscopes to extraordinary
accuracy.  Such precessions are predicted by General Relativity (GR), and one component of this precession is 
the `frame-dragging'  or Lense-Thirring effect, which is caused by the rotation of the earth, and the other is  
the geodetic effect.   A new theory of gravity predicts, however,  
a second and much larger `frame-dragging' or vorticity induced spin precession.  This  spin precession component will 
also display the effects of novel gravitational waves which are predicted by the new theory of gravity, and 
which have already been seen in several experiments. The magnitude and signature of these gravitational wave induced 
spin precession effects is given for comparison with the GP-B experimental data.

\newpage
\tableofcontents

\section{ Introduction}
 The Stanford University-NASA Gravity Probe B satellite experiment has the
 capacity to measure the precession of four on-board gyroscopes to unprecedented
accuracy \cite{Pugh,Schiff,PEV,Everitt,GPB}.  The experiment was proposed independently by 
George Pugh in 1959 and Leonard Schiff in 1960.  Such a
precession is predicted by the Einstein theory of gravity, General Relativity (GR), with
 two components: (i) a geodetic precession, and (ii) a `frame-dragging' precession
known as the Lense-Thirring effect.  The latter is a particularly interesting effect
 induced by the rotation of the earth, and described in GR in terms of a
`gravitomagnetic' field.  According to GR this smaller effect will give
a precession perpendicular to the plane of the satellite orbit  accumulating to   0.042 arcsec per year for the
GP-B gyroscopes.  However a recently developed  theory  gives a different account of gravity \cite{NovaDM,NovaBook}.
This theory gives a dynamical account of the so-called
`dark matter' effect in spiral galaxies. It also successfully predicts the masses of the black holes found in the
globular clusters M15 and G1.  Here we show that GR and the new theory make very
different predictions for the `frame-dragging' effect, and so the GP-B experiment will
be able to decisively test both theories. While predicting the same earth-rotation
induced precession, the new theory has an additional much larger `frame-dragging'
effect caused by the observed translational motion of the earth, and in a different direction to the earth
induced rotation induced precession.  As well the new non-metric theory explains the `frame-dragging' effect in
terms of  vorticity in a `substratum flow'.  This `flow' exhibits fluctuations or wave effects that have already
been seen in at least three experiments. These are the gravitational waves of the new theory of gravity, and are
completely different from the gravitational waves predicted by General Relativity.
 Herein the magnitude and signature of this new component of the  gyroscope
precession is predicted, with particular emphasis on the gravitational wave effects, which it is predicted
will be detectable by the GP-B experiment.

\section{ Aspects of the  History of Gravity Research}

The history of the development of theories of gravity is well known.   Newton introduced the
gravitational acceleration field ${\bf g}({\bf r},t)$, whose source was matter, as shown in (\ref{eqn:NG}),
and which was then used by Newton to explain the solar system planetary dynamics as had been expressed by
Kepler in his laws.   In terms of this acceleration field Newton's `universal law of gravitation', in (\ref{eqn:Newton}),
is uniquely determined by Kepler's laws. However the phenomena of gravity may be equally well represented in terms of a
`flow' system involving a velocity vector field ${\bf v}({\bf r},t)$, and this formalism is physically
indistinguishable from the Newtonian formalism, in terms of the consequent gravitational forces in the solar system, though
it is clearly different mathematically.  However, unlike the Newtonian formalism, this velocity field formalism permits an
additional dynamical effect that would not have manifested in the case of the solar system, that is, it is not
inconsistent with Kepler's laws. It is now known that this additional dynamical effect is nothing more than the
so-called `dark matter' effect \cite{NovaDM}.  So we now understand that had Newton used the velocity field instead of the
acceleration field as the fundamental degrees of freedom then the `dark matter' effect could have been
predicted by Newton long ago.

There was another `accident of history' that has enormously influenced the development of physics, and in particular
that of gravity, namely the detection of absolute motion.  In this case Michelson and Morley in their interferometer
experiment of 1887 detected the expected fringe shifts caused by absolute motion, but using Newtonian physics Michelson had
developed a theory for the instrument which predicted fringe shifts larger than actually observed.  This
supposed failure to detect the expected fringe shifts and so absolute motion led to the notion that absolute
motion, that is, motion measured with respect to local space itself, is not even a meaningful concept,
despite that over the last 114 years it has been detected in at least seven experiments, and of several
kinds.  Re-analysing the operation of the Michelson interferometer by taking account, for the first time, of
both the Fitzgerald-Lorentz contraction effect and the effect of a gas being present in the light path, we
now know that the Michelson-Morley experiment gives a lower limit of some 300 km/s for the speed of
absolute motion of the earth through space\footnote{A lower limit arises because of the projection effect
occurring in interferometers.}, rather than the 8 km/s which follows from the Newtonian theory and which they
reported in 1887. The existence of absolute motion and of the gravitational effects associated with that  `flow'
have gone unstudied for the last 100 years, and the `dark-matter' effect is only one of numerous gravitational
phenomena that have been observed but remain as unexplained - these phenomena are known as `gravitational
anomalies'.  It always needs to be emphasised that absolute motion is the cause of the various relativistic
effects, an idea that goes back to Lorentz in the 19$^{th}$ century.

In 1916 Hilbert and Einstein proposed a metric spacetime theory of gravity,    which was constructed to agree with
the Newtonian acceleration field formalism in the so-called non-relativistic limit. But in doing so Hilbert
and Einstein consequently missed the `dark matter' effect, as that effect had been missed also by Newton.  As
well  this theory also missed the existence of absolute  motion, an effect absolutely critical to
understanding gravity.  It has been repeatedly claimed that the Hilbert-Einstein General Theory of Relativity has
been confirmed many times, but this is untrue.  All but one of the so-called tests merely used the
geodesic equation which determines the trajectory of a particle or an electromagnetic wave  in a given metric.  That
metric has in all cases been the external Schwarzschild metric, but apparently unknown to most is that this metric
is nothing more than the Newtonian `inverse square law' in mathematical disguise, namely with the metric
expressed in terms of the particular velocity vector flow field corresponding to Newton's inverse square law.  So
these tests of GR were confirming, at best,  the flow formalism for gravity, together with its geodesic equation,
and had nothing to do with the dynamical content of GR.  Indeed the one single test, so far, of GR is the observed
decay rate of the binary pulsar orbits. As well, as already noted, GR has failed  in the case of the  `dark matter'
effect. 

The Gravity Probe B experiment will test the dynamics of GR and of the new theory, and the differences
between the two predictions are very different.  Because of the denial of absolute linear motion in GR it
predicts only a `frame-dragging' effect  produced by the absolute rotation of the earth, whereas the new
theory of gravity predicts not only that effect, but most significantly and dominantly, a `frame-dragging'
effect  caused by the absolute linear motion of the earth.  In the new theory, because it is not a metric
spacetime theory, the `spacetime frame-dragging' is seen to be a vorticity in the flow of space caused by the
absolute  motion of the earth, both rotational and linear, through that space.  As well that flow has wave
phenomena which can be detected via their effect on the vorticity.  The vorticity of the flow, which is
simply a local  rotation of the direction of the flow, affects the GP-B gyroscopes, and anything else, by
simply rotating the whole system.   So  the gyroscope spin precessions can detect these gravitational waves. 

As well the `dark matter' dynamical effect has its magnitude determined by a new dimensionless gravitational
constant, in addition to the Newtonian constant $G$. By comparing the new theory with the experimental data
from the Greenland borehole $g$ anomaly data, it was found that the new gravitational constant is the fine
structure constant $\alpha \approx 1/137$. This has enormous implications for fundamental physics, as
discussed elsewhere \cite{NovaDM}, but in particular the new theory predicts black holes whose dynamics is
determined by the value of $\alpha$, and not by $G$. This result has been confirmed by  the  black holes observed
in several globular clusters. The `dark matter' effect is also apparent in Cavendish-type experiments that 
measure $G$, and there has been a longstanding unknown systematic effect present in these experiments that has caused
$G$ to be the least accurately known fundamental constant \cite{NovaDM}.

\section{ A Theory of Gravity}

Here we briefly review the derivation of the new theory of gravity \cite{NovaDM,NovaBook}. The Newtonian
`inverse square law' for gravity,
\begin{equation}
F=\frac{Gm_1m_2}{r^2},
\label{eqn:Newton}\end{equation}   was based on Kepler's laws for the motion of the planets.  Newton's 
`explanation' of the phenomena of gravity was in terms of  the gravitational acceleration vector field
${\bf g}({\bf r},t)$, and in differential form 
\begin{equation}\label{eqn:NG}
\nabla.{\bf g}=-4\pi G\rho,
\end{equation}
where $\rho({\bf r},t)$ is the matter density.  However there is an alternative formulation \cite{NovaDM} in terms of
 a vector `flow' field ${\bf v}({\bf r},t)$ determined by
\begin{equation}
\frac{\partial }{\partial t}(\nabla.{\bf v})+\nabla.(({\bf
v}.{\bf \nabla}){\bf v})=-4\pi G\rho,
\label{eqn:CG1}\end{equation}
with ${\bf g}$ now given by the Euler `fluid' acceleration
\begin{equation}{\bf g}=\displaystyle{\frac{\partial {\bf v}}{\partial
t}}+({\bf v}.{\bf
\nabla}){\bf v}=\displaystyle{\frac{d{\bf v}}{dt}}.
\label{eqn:CG2}\end{equation}
Trivially this ${\bf g}$ also satisfies (\ref{eqn:NG}).
External to a spherical mass $M$ of radius $R$ a velocity field solution of (\ref{eqn:CG1}) is 
\begin{equation}
{\bf v}({\bf r})=-\sqrt{\frac{2GM}{r}}\hat{\bf r},  \mbox{\ \ }r>R,
\label{eqn:vfield}\end{equation}
which gives from (\ref{eqn:CG2}) the usual inverse square law ${\bf g}$ field
\begin{equation}
{\bf g}({\bf r})=-\frac{GM}{r^2}\hat{\bf r}, \mbox{\ \ }r>R.
\label{eqn:InverseSqLaw}\end{equation} 
\index{inverse square law}
However the flow equation (\ref{eqn:CG1}) is not uniquely determined by Kepler's laws because  
\begin{equation}
\frac{\partial }{\partial t}(\nabla.{\bf v})+\nabla.(({\bf
v}.{\bf \nabla}){\bf v})+C({\bf v})=-4\pi G\rho,
\label{eqn:CG3}\end{equation}
where
\begin{equation}
C({\bf v})=\displaystyle{\frac{\alpha}{8}}((tr D)^2-tr(D^2)),
\label{eqn:Cdefn1}\end{equation} and
\begin{equation}
D_{ij}=\frac{1}{2}\left(\frac{\partial v_i}{\partial x_j}+\frac{\partial v_j}{\partial x_i}\right),
\label{eqn:Ddefn1}\end{equation}
also has the same external solution (\ref{eqn:vfield}),  because $C({\bf v})=0$ for the flow in 
(\ref{eqn:vfield}). So the presence of the  $C({\bf v})$ would not have manifested in the special case
of planets in orbit about the massive central sun.
Here $\alpha$ is a   dimensionless constant - a new gravitational constant, in addition to usual
the Newtonian gravitational constant $G$. However inside a spherical mass we find \cite{NovaDM} that 
$C({\bf v})\neq 0$, and using the Greenland  borehole $g$ anomaly data \cite{Greenland} we find that
$\alpha^{-1}=139 \pm  5 $, which gives the fine structure constant $\alpha=e^2\hbar/c \approx 1/137$
to within experimental error. From (\ref{eqn:CG2}) we can write
\begin{equation}\label{eqn:g2}
\nabla.{\bf g}=-4\pi G\rho-4\pi G \rho_{DM},
\end{equation}
where
\begin{equation}
\rho_{DM}({\bf r})=\frac{\alpha}{32\pi G}( (tr D)^2-tr(D^2)),  
\label{eqn:DMdensity}\end{equation} 
which introduces an effective `matter density' representing the flow dynamics associated with the 
$C({\bf v})$ term. In \cite{NovaDM} this dynamical effect is shown to be the `dark matter' effect.
The interpretation of the vector flow field ${\bf v}$ is that it is a manifestation, at the classical
level, of a quantum substratum to space; the flow is a rearrangement of that substratum, and not a flow
{\it through}  space. However  (\ref{eqn:CG3}) needs to be further generalised \cite{NovaDM} to include
vorticity, and also the effect of the motion of matter through this substratum via  
\begin{equation}
{\bf v}_R({\bf r}_0(t),t) ={\bf v}_0(t) - {\bf v}({\bf r}_0(t),t),
\label{eqn:CG$}
\end{equation}
where ${\bf v}_0(t)$ is the velocity of an object, at ${\bf r}_0(t)$, relative to the same frame of reference that defines
the flow field; then ${\bf v}_R$ is the velocity of that matter relative to the substratum. The flow equation
(\ref{eqn:CG3}) is then generalised to, with $d/dt=\partial/\partial t +{\bf v}.\nabla$  the Euler fluid or total
derivative,
\begin{eqnarray}
&&\frac{d D_{ij}}{dt}+ \frac{\delta_{ij}}{3}tr(D^2) + \frac{tr D}{2}
(D_{ij}-\frac{\delta_{ij}}{3}tr D)+\frac{\delta_{ij}}{3}\frac{\alpha}{8}((tr
D)^2 -tr(D^2))\nonumber \\ && +(\Omega D-D\Omega)_{ij}=-4\pi
G\rho(\frac{\delta_{ij}}{3}+\frac{v^i_{R}v^j_{R}}{2c^2}+..),\mbox{ } i,j=1,2,3. 
\label{eqn:CG4a}\end{eqnarray}
\begin{equation}\nabla \times(\nabla\times {\bf v}) =\frac{8\pi G\rho}{c^2}{\bf v}_R,
\label{eqn:CG4b}\end{equation}
\begin{equation}
\Omega_{ij}=\frac{1}{2}\left(\frac{\partial v_i}{\partial x_j}-\frac{\partial v_j}{\partial
x_i}\right)=-\frac{1}{2}\epsilon_{ijk}\omega_k=-\frac{1}{2}\epsilon_{ijk}(\nabla\times {\bf v})_k,
\label{eqn:BS}\end{equation}
and the vorticity vector field is $\vec{\omega}=\nabla\times {\bf v}$. For zero vorticity and
$v_R\ll c$ (\ref{eqn:CG4a}) reduces to (\ref{eqn:CG3}).   We obtain from
(\ref{eqn:CG4b}) the Biot-Savart form for the vorticity 
\begin{equation}
\vec{\omega}({\bf r},t)
=\frac{2G}{c^2}\int d^3 r^\prime \frac{\rho({\bf r}^\prime,t)}
{|{\bf r}-{\bf r}^\prime|^3}{\bf v}_R({\bf r}^\prime,t)\times({\bf r}-{\bf r}^\prime).
\label{eqn:omega}\end{equation} 

The GP-B experiment will detect this vorticity field by means of the spin-axis rotation or precession of the
gyroscopes over time, with the magnitude and direction of this rotation or precession measured relative to the
direction of the guide star as observed by the on-board telescope. Clearly any wave phenomena that changes ${\bf
v}_R({\bf r},t)$  in (\ref{eqn:omega}) will also be detected via its effect on $\vec{\omega}({\bf r},t)$.

\section{ Geodesics}

We now define how the trajectory of a point object is determined by the  velocity flow field, which manifests via
the velocity of absolute motion of the object relative to the local space, namely via ${\bf v}_R$. There is a
problem with terminology here: what is called absolute motion here is actually motion with respect to the local
substratum structure of space, which means that it is a relative motion. However in the language and restrictions of
conventional `special relativity' all velocities are relative, where in this case it means relative to another
object and {\it not} relative to space itself.  Most significantly the geodesic equation herein involves {\it both}
(absolute) motion with respect to space  and also the relativistic time dilation effect. This latter effect
involves the notion that absolute motion, whether linear or rotational,  causes, say,  a clock moving through space
to tick more slowly than one at rest in space. Similarly in the re-analysis of the principles of operation of the
Michelson interferometer it was necessary to take account of both absolute linear motion {\it and } the
relativistic length contraction  effect upon the arms of the interferometer caused by that absolute linear motion. 
So again to observe absolute motion we must take account of those  relativistic effects which are actually caused
by absolute motion. This is contrary to the postulate by Einstein which asserts that absolute motion has no meaning
and so no experimental manifestation.  The GP-B experiment will, yet again, show that this key postulate is invalid, but
which does not invalidate the phenomena known as `special relativistic' effects. In terms of the history of physics it
implies that we must return to the pre-Einstein ideas of Lorentz and others.

The path
${\bf r}_0(t)$ of an object through space  is obtained by  extremising  the relativistic proper time
\begin{equation}
\tau[{\bf r}_0]=\int dt \left(1-\frac{{\bf v}_R^2}{c^2}\right)^{1/2}
\label{eqn:CG5}\end{equation}
This entails the idea that the speed of light $c$ is the maximum speed through the local space.  This means that
the speed of light is $c$ only with respect to the local space. That  it is believed that $c$ is the speed of light
for all observers in uniform linear motion, as postulated by Einstein,  is an error that follows from not realising
that when in motion the observer's clock and rod are affected by that motion. Without correcting for such absolute
motion effects   the incorrect notion of $c$ being the `universal speed of light' is not realised. 

To extremise $\tau$ we use a small  deformation of the trajectory 
\begin{equation}{\bf r}_0(t)
\rightarrow  {\bf r}_0(t) +\delta{\bf r}_0(t) \mbox{\  \ giving \  \ }
{\bf v}_0(t) \rightarrow  {\bf v}_0(t) +\displaystyle\frac{d\delta{\bf r}_0(t)}{dt},\end{equation}
  and then we also
have
\begin{equation}\label{eqn:G2}
{\bf v}({\bf r}_0(t)+\delta{\bf r}_0(t),t) ={\bf v}({\bf r}_0(t),t)+(\delta{\bf
r}_0(t).{\bf \nabla}) {\bf v}({\bf r}_0(t))+... 
\end{equation}
Then
\begin{eqnarray}\label{eqn:G3}
\delta\tau&=&\tau[{\bf r}_0+\delta{\bf r}_0]-\tau[{\bf r}_0]  \nonumber\\
&=&-\int dt \:\frac{1}{c^2}{\bf v}_{{R}}. \delta{\bf v}_{{R}}\left(1-\displaystyle{\frac{{\bf
v}_{{R}}^2}{c^2}}\right)^{-1/2}+...\nonumber\\
&=&\int dt\frac{1}{c^2}\left({\bf
v}_{{R}}.(\delta{\bf r}_0.{\bf \nabla}){\bf v}-{\bf v}_{{R}}.\frac{d(\delta{\bf
r}_0)}{dt}\right)\left(1-\displaystyle{\frac{{\bf v}_{{R}}^2}{c^2}}\right)^{-1/2}+...\nonumber\\ 
&=&\int dt \frac{1}{c^2}\left(\frac{{\bf v}_{{R}}.(\delta{\bf r}_0.{\bf \nabla}){\bf v}}{ 
\sqrt{1-\displaystyle{\frac{{\bf
v}_{{R}}^2}{c^2}}}}  +\delta{\bf r}_0.\frac{d}{dt} 
\frac{{\bf v}_{{R}}}{\sqrt{1-\displaystyle{\frac{{\bf
v}_{{R}}^2}{c^2}}}}\right)+...\nonumber\\
&=&\int dt\: \frac{1}{c^2}\delta{\bf r}_0\:.\left(\frac{({\bf v}_{{R}}.{\bf \nabla}){\bf v}+{\bf
v}_{{R}}\times({\bf
\nabla}\times{\bf v})}{ 
\sqrt{1-\displaystyle{\frac{{\bf
v}_{{R}}^2}{c^2}}}}  +\frac{d}{dt} 
\frac{{\bf v}_{{R}}}{\sqrt{1-\displaystyle{\frac{{\bf
v}_{{R}}^2}{c^2}}}}\right)+...\nonumber\\
& &
\end{eqnarray}
  Hence a 
trajectory ${\bf r}_0(t)$ determined by $\delta \tau=0$ to $O(\delta{\bf r}_0(t)^2)$ satisfies 
\begin{equation}\label{eqn:G4}
\frac{d}{dt} 
\frac{{\bf v}_{{R}}}{\sqrt{1-\displaystyle{\frac{{\bf v}_{{R}}^2}{c^2}}}}=-\frac{({\bf
v}_{{R}}.{\bf \nabla}){\bf v}+{\bf v}_{{R}}\times({\bf
\nabla}\times{\bf v})}{ 
\sqrt{1-\displaystyle{\frac{{\bf v}_{R}^2}{c^2}}}}.
\end{equation}
  Substituting ${\bf
v}_{{R}}(t)={\bf v}_0(t)-{\bf v}({\bf r}_0(t),t)$ and using 
\begin{equation}\label{eqn:G5}
\frac{d{\bf v}({\bf r}_0(t),t)}{dt}=({\bf v}_0.{\bf \nabla}){\bf
v}+\frac{\partial {\bf v}}{\partial t},
\end{equation}
and then
\begin{equation}\label{eqn:CG6}
\frac{d}{dt} 
\frac{{\bf v}_0}{\sqrt{1-\displaystyle{\frac{{\bf v}_R^2}{c^2}}}}={\bf v}
\frac{d}{dt}\frac{1}{\sqrt{1-\displaystyle{\frac{{\bf v}_R^2}{c^2}}}}+\frac{\displaystyle{\frac{\partial {\bf
v}}{\partial t}}+({\bf v}.{\bf \nabla}){\bf v}+({\bf \nabla}\times{\bf v})\times{\bf v}_R}{ 
\displaystyle{\sqrt{1-\frac{{\bf v}_R^2}{c^2}}}},
\end{equation}
and  finally 
\begin{equation}\label{eqn:newaccel}
 \frac{d {\bf v}_0}{dt}=-\frac{{\bf
v}_R}{1-\displaystyle{\frac{{\bf v}_R^2}{c^2}}}
\frac{1}{2}\frac{d}{dt}\left(\frac{{\bf v}_R^2}{c^2}\right)
+\left(\displaystyle{\frac{\partial {\bf v}}{\partial t}}+({\bf v}.{\bf \nabla}){\bf
v}\right)+({\bf
\nabla}\times{\bf v})\times{\bf v}_R.
\end{equation}
This is a generalisation of the acceleration in (\ref{eqn:CG2}) to include the vorticity effect, as the last term,
and the first term which  is the resistance to acceleration caused by the relativistic `mass' increase effect.
This term leads to the so-called geodetic effects.
  The vorticity term causes the GP-B gyroscopes to develop the vorticity induced precession
\cite{GPBCahill}, which is simply the rotation of space carrying the gyroscope along with it, compared to
more distant space which is not involved in that rotation.   The middle
term, namely the acceleration in (\ref{eqn:CG2}), is simply the usual Newtonian gravitational acceleration,
but now seen to arise from the inhomogeneity and time-variation of the flow velocity field. As already noted it was
this geodesic equation that has been checked in various experiments, but always, except in the case of the binary
pulsar slow-down, with the velocity field given by  the Newtonian `inverse square law' equivalent form in
(\ref{eqn:vfield}). As discussed elsewhere \cite{NovaDM,NovaBook} this flow is exactly equivalent to the external
Schwarzschild metric. 

\section[ Gravitational Waves]{ Gravitational Waves \label{section:waves}}
Newtonian gravity in its original `force' formalism (\ref{eqn:NG}) does not admit any wave phenomena.
However the  `in-flow' formalism  (\ref{eqn:CG4a})-(\ref{eqn:CG4b}) does admit  wave
phenomena. For the simpler  case here of very small  vorticity and no  motion of matter effects, namely  ${\bf v}_R^2
\ll c^2$ on the RHS of  (\ref{eqn:CG4a}),  (\ref{eqn:CG4a})-(\ref{eqn:CG4b}) reduce to
(\ref{eqn:CG1}). This is seen by writing ${ \bf v}=\nabla u$, where $u$ is  the velocity potential, which is valid when
$\nabla \times {\bf v}=0$.
 Here we shall  also neglect the `dark matter' effect.
Then in terms of $u({\bf r},t)$ (\ref{eqn:CG1}) becomes 
\index{velocity potential}
\begin{equation}
\frac{\partial  u}{\partial t}+\frac{1}{2}(\nabla u)^2=-\Phi.
\label{eqn:NGu}\end{equation}
with
\begin{equation}{\bf g}=\frac{ \partial \nabla u}{\partial t}+\frac{1}{2}\nabla(\nabla u)^2, 
\label{eqn:gu}\end{equation} 
where $\Phi$ is the Newtonian gravitational potential
\begin{equation}\label{eqn:Phieqn}
\Phi({\bf r},t)=-G\int d^3 r^\prime\frac{\rho({\bf r}^\prime,t)}{|{\bf r}-{\bf r}^\prime|}.
\end{equation}
Equations (\ref{eqn:NGu}) and (\ref{eqn:gu})  together exactly  reproduce (\ref{eqn:NG}), even when the flow is
time-dependent, and whether or not the matter density is time dependent. We note that the `inverse square
law', in the `flow' formalism, follows from the structure of the LHS of (\ref{eqn:CG1}), which in turn is
determined by the Galilean covariance of the Euler `fluid' derivative \cite{Super}. In this regard we note that the
Galilean covariance is not at odds with  Lorentz covariance, provided that we understand that they
apply to
 observer data either after or before correcting the data for the effects of absolute motion upon the observers
instruments \cite{NovaBook}.

We shall now show that in terms of the velocity field formalism  non-relativistic  gravity possess a
wave phenomena.  So again had Newton used this velocity formalism then he could have predicted the existence of
such gravitational waves.  These gravitational waves are totally different from those predicted by General
Relativity, and unlike these, which have not been observed, the new gravitational waves can now be understood to 
have been seen in numerous experiments, as discussed later and in \cite{RGC}.

Suppose that (\ref{eqn:NGu}) has  for a static matter density a static solution $u_0({\bf r})$ with
corresponding velocity field
${\bf v}_0({\bf r})$, and with corresponding acceleration
${\bf g}_0(\bf r)$. Then we look for time dependent perturbative solutions of (\ref{eqn:NGu}) with
$u=u_0+ \overline{u}$. To first order in $\overline{u}$ we then have
\begin{equation}
\frac{\partial {\overline u({\bf r},t)}}{\partial t}=-{\bf \nabla}\overline{u}({\bf r},t).{\bf
\nabla}u_0({\bf r}),
\label{eqn:ueqn3}\end{equation}
This equation is easily seen to have wave solutions of the form $\overline{u}({\bf r},t)=A\cos({\bf k}.{\bf
r}-\omega t+\phi)$ where $\omega({\bf k},{\bf r})={\bf v}_0({\bf r}).{\bf k}$, for wavelengths short compared to 
the scale of changes in  ${\bf v}_0({\bf r})$. The phase velocity of these waves is then
${\bf v}_\phi={\bf v}_0$, and the group velocity is ${\bf v}_g={\bf \nabla}_k\omega={\bf 
v}_0$. Then the velocity field is 
\begin{equation}
{\bf v}({\bf r},t)={\bf v}_0({\bf r})-A{\bf k}\sin({\bf k}.{\bf r}-w({\bf k},{\bf r})t+\phi).
\end{equation}
In general we have, perturbatively, the superposition of such waves, giving
\begin{equation}
{\bf v}({\bf r},t)={\bf v}_0({\bf r})-\int d^3k
A({\bf k}){\bf k}\sin({\bf k}.{\bf r}-w({\bf k},{\bf
r})t+\phi({\bf k})).
\end{equation}
The wave part of this expression, it is suggested, describes the wave phenomena shown in
Fig.\ref{fig:Turbulence}. These wave solutions have also been seen in non-perturbative
numerical solutions of (\ref{eqn:CG1}), and even when the `dark matter' effect is retained as
in (\ref{eqn:CG3}).

But are these wave solutions physical, or are they a mere artifact of
the in-flow formalism?  First note that the wave phenomena do not cause any gravitational effects, 
in the above approximation, because the acceleration field is independent of their existence;  whether they
are present or not does not affect
 ${\bf g}({\bf r})$.  Hence  these gravitational waves produce no force
effects via the acceleration field defined by (\ref{eqn:CG2}).  The question is equivalent to
asking which of the fields ${\bf v}$ or ${\bf g}$ is the fundamental quantity.  As we have already
noted  the velocity field ${\bf v}$  and these wave phenomena have already  been observed
\cite{RGC,AMGE,NovaBook}. Indeed it is even possible that the effects of such waves are present in
the Michelson-Morley 1887 fringe shift data.  This would imply that the real gravitational waves
have actually been observed for over 100 years. 

Within the full new theory of gravity these
waves do affect the acceleration field ${\bf g}$, via the dynamical influence of the new $C({\bf v})$ `dark
matter' term, and also via the goedesic  and  vorticity terms in (\ref{eqn:newaccel}).  There is 
evidence that the effects of these two terms have been seen in the experiments by Allais, Saxl and Allen, and
by Zhou \cite{AMGE}.   The observational evidence is that these gravitational waves are in the main associated
with gravitational phenomena in the Milky Way and local galactic cluster, as revealed in the analysis of data
from at least three distinct observations of absolute motion effects
\cite{AMGE}.

\section{ Vorticity Effects} 

\begin{figure}[t]
\hspace{35mm}\includegraphics[scale=1.0]{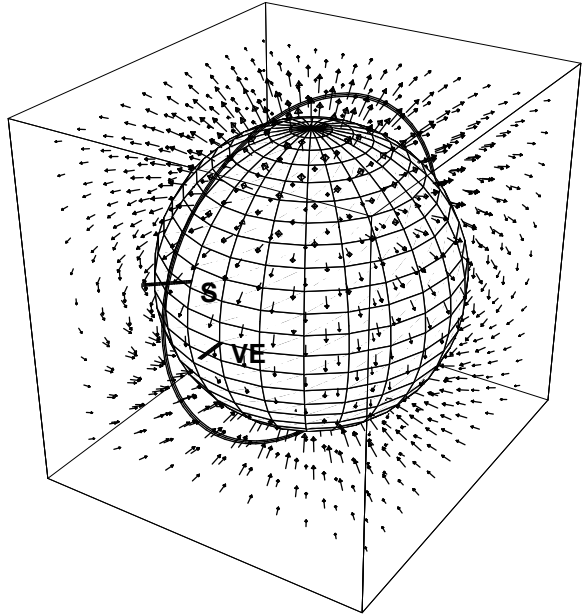}
\caption{\small{ Shows the earth (N is up) and vorticity vector field component $\vec{\omega}$ induced by the rotation of the
earth, as in  (\ref{eqn:rotation}). The polar orbit of the GP-B satellite is shown,    ${\bf S}$ is the gyroscope starting 
spin orientation, directed towards the guide  star IM Pegasi, RA = $22^h $ $53^\prime$ $ 2.26^{\prime\prime}$, Dec = $16^0$ $
50^\prime $ $28.2^{\prime\prime}$, and  ${\bf VE}$ is the vernal equinox.}
\label{fig:Rotation}}\end{figure}

\begin{figure}[t]
\hspace{35mm}\includegraphics[scale=1.0]{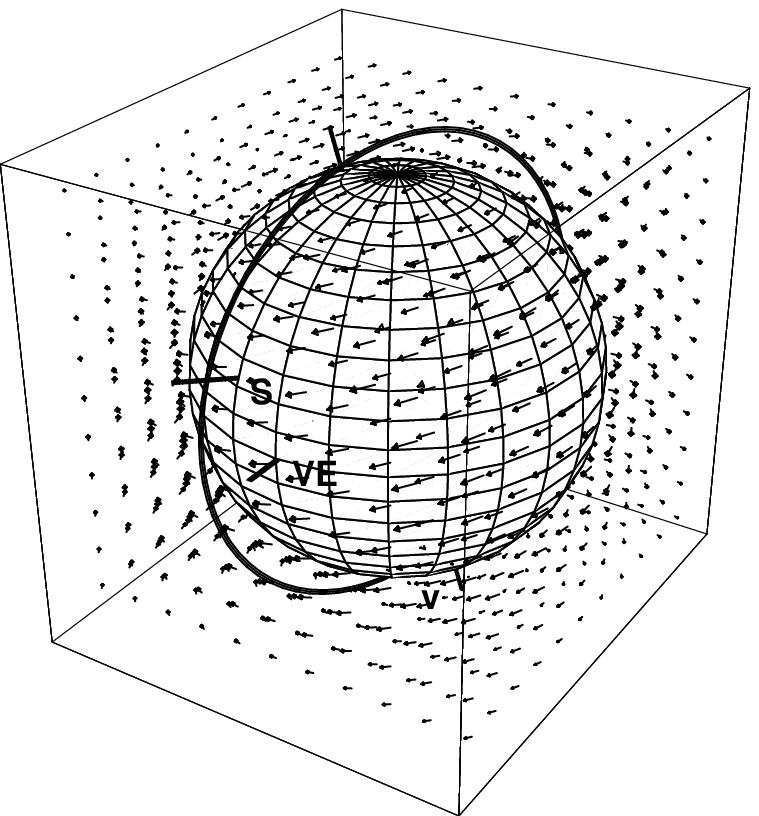}
\caption{\small{ Shows the earth (N is up) and the much larger vorticity vector field component $\vec{\omega}$ induced by the
translation of the earth, as in  (\ref{eqn:AMomega}). The polar orbit of the GP-B satellite is shown, and   ${\bf S}$ is the
gyroscope starting  spin orientation, directed towards the guide  star IM Pegasi,  RA = $22^h $ $53^\prime$ $ 2.26^{\prime\prime}$, Dec
= $16^0$ $ 50^\prime $ $28.2^{\prime\prime}$,   ${\bf VE}$ is the vernal equinox,
and ${\bf V}$ is the direction  $\mbox{RA} = 5.2^{h}$, $\mbox{Dec} = -67^0$ of the translational velocity ${\bf v}_c$.}
\label{fig:Absolute}}\end{figure}

Here we consider one difference between the two theories, namely that associated with the vorticity
part of  (\ref{eqn:newaccel}), leading to the `frame-dragging' or Lense-Thirring  effect. In GR the vorticity field
 is known as the `gravitomagnetic' field ${\bf B}=-c\:\vec{\omega}$. In  both GR and  the new
theory the vorticity is given by (\ref{eqn:omega}) but with a key difference: in GR ${\bf v}_R$ is {\it only} the
rotational velocity of the matter in the earth, whereas in (\ref{eqn:CG4a})-(\ref{eqn:CG4b})
${\bf v}_R$ is the vector sum of the  rotational velocity and the translational velocity of the
earth through the substratum.   At least seven experiments have detected this translational velocity; some
were gas-mode Michelson interferometers and others coaxial cable experiments \cite{RGC,AMGE,NovaBook}, and the
translational velocity is now known to be  approximately  430 km/s in the direction RA $ = 5.2^{h}$,
Dec$ = -67^0$. This direction has been known since  the Miller
\cite{Miller} gas-mode interferometer experiment, but the RA was more recently confirmed by the 1991
 DeWitte coaxial cable experiment performed in the Brussels laboratories of Belgacom \cite{AMGE}. 
This flow is related to galactic gravity flow effects  \cite{RGC,AMGE,NovaBook}, and so  is different to that
of the velocity of the earth with respect to the Cosmic Microwave Background (CMB), which is $369$ km/s in the direction
 $\mbox{RA }=11.20^h,\mbox{Dec } =-7.22^0$.

First consider the common but much smaller rotation induced `frame-dragging' or vorticity effect. Then
${\bf v}_R({\bf r})={\bf w}\times{\bf r}$ in (\ref{eqn:omega}), where ${\bf w}$ is the angular
velocity of the earth, giving
\begin{equation}
\vec{\omega}({\bf r})=4\frac{G}{c^2}\frac{3({\bf r}.{\bf L}){\bf r}-r^2{\bf L}}{2 r^5},
\label{eqn:rotation}\end{equation}
where ${\bf L}$ is the \index{angular momentum - earth} angular momentum of the earth, and ${\bf
r}$ is the distance from the centre. This component of the vorticity field is shown in
Fig.\ref{fig:Rotation}.  Vorticity may be detected by observing the precession of the GP-B
gyroscopes.  The vorticity term in 
 (\ref{eqn:CG6}) leads to a torque on the angular momentum ${\bf S}$ of the gyroscope,
\begin{equation}
\vec{\tau}= \int d^3 r \rho({\bf r})\; {\bf r}\times(\vec{\omega}({\bf r}) \times{\bf v}_R({\bf r})),
\label{eqn:torque1}\end{equation}
where $\rho$ is its  density, and where
  ${\bf v}_R$ is used here to describe the rotation of the gyroscope.  Then $d{\bf S}=\vec{\tau}dt$ is the change in
${\bf S}$ over the time interval $dt$. In the above case 
${\bf v}_R({\bf r})={\bf s}\times{\bf r}$, where ${\bf s}$ is the angular velocity of the gyroscope.  
This gives
\begin{equation}
\vec{\tau}=\frac{1}{2}\vec{\omega}\times{\bf S}
\label{eqn:torque2}\end{equation}
and so $\vec{\omega}/2$ is the instantaneous angular velocity of precession of the gyroscope. This corresponds to
the well known fluid result that the vorticity vector is twice the angular velocity vector.   For GP-B the direction
of
${\bf S}$   has been chosen so that this precession is cumulative and, on averaging  over an orbit,
corresponds to some $7.7\times 10^{-6}$ arcsec per orbit, or 0.042 arcsec per year.  GP-B has been superbly
engineered so that measurements to a precision of 0.0005 arcsec are possible. 

\begin{figure}[t]
\hspace{10mm}\includegraphics[scale=1.5]{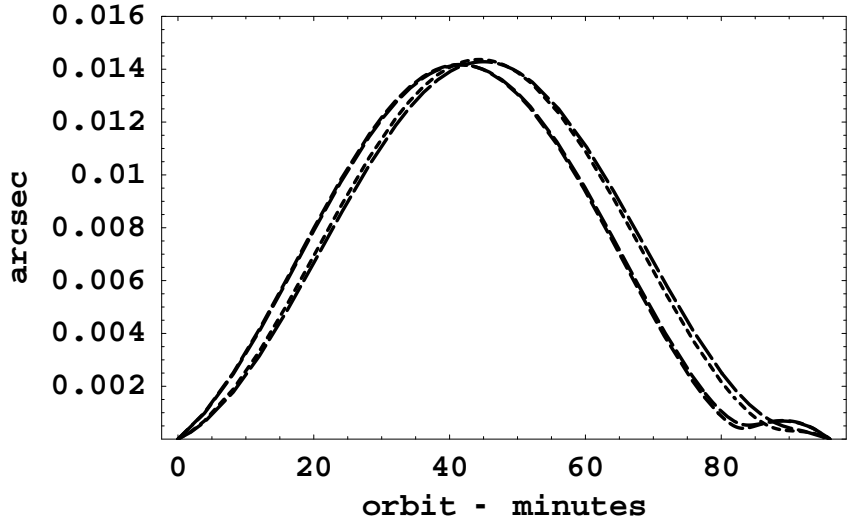}
\caption{\small{ Predicted  variation of the precession angle  $\Delta \Theta=|\Delta {{\bf S}}(t)|/|{\bf S}(0)|$, in
arcsec, over one 97 minute GP-B orbit, from the vorticity induced by the translation of the earth, as given by
(\ref{eqn:precession}). The orbit time begins at location
${\bf S}$. Predictions are for the  months of April, August, September and February, labeled by  increasing dash
length.  The `glitches' near 80 minutes are caused by the angle effects in (\ref{eqn:precession}). These changes arise
from the effects of the changing orbital velocity of the earth about the sun.  The GP-B expected angle measurement
accuracy is 0.0005 arcsec. The gravitational waves will  affect these plots, as shown in Fig.\ref{fig:waveprecessions}
for 40 orbits.}
\label{fig:Precession}}\end{figure}

\begin{figure}[h]
\hspace{35mm}\includegraphics[scale=1.0]{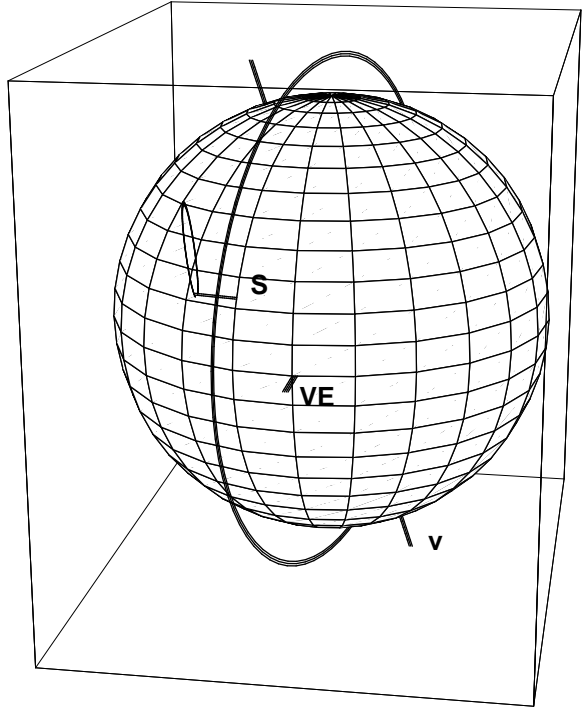}
\caption{\small{ Predicted precession of the gyroscope spin axis, over one
orbit of the satellite, but greatly exaggerated,  directly manifesting the
 vorticity component of the flow caused by the translation of the earth, as
in  (\ref{eqn:AMomega}) and Fig.\ref{fig:Absolute}. This component of the spin precession
forms an elongated ellipse. Gravitational waves will cause changes in the size and orientation
 of this precession ellipse. The angle $\Delta
\Theta$ in Fig.\ref{fig:Precession} is  the angle subtended at the earth's centre by the starting position at ${\bf S}$
and a point on this ellipse of precession. The  precession caused by the vorticity component arising
from the rotation of the earth is perpendicular to the plane of the orbit, while the
geodetic precession component is in the plane of the orbit.  The polar orbit
of the GP-B satellite is shown, and  
${\bf S}$ is the gyroscope starting  spin orientation, directed towards the guide  star IM
Pegasi,  RA =
$22^h $ $53^\prime$ $ 2.26^{\prime\prime}$, Dec = $16^0$ $ 50^\prime
$ $28.2^{\prime\prime}$,   ${\bf VE}$ is the vernal equinox, and ${\bf V}$ is the
direction  $\mbox{RA} = 5.2^{h}$, $\mbox{Dec} = -67^0$ of the translational velocity ${\bf
v}_c$.}
\label{fig:PrecGraphic}}\end{figure}

However for the unique translation-induced precession if we  use $v_R \approx v_C = 430$ km/s in the
direction  $\mbox{RA} =5.2^{hr}$, $\mbox{Dec} =-67^0$, namely ignoring the effects of the orbital motion of the
earth, the observed flow past the earth towards the sun, and the flow into the earth, and effects of
the gravitational waves, then (\ref{eqn:omega}) gives
\begin{equation}
\vec{\omega}({\bf r})=\frac{2GM}{c^2}\frac{{\bf v}_C\times{\bf r}}{r^3}.
\label{eqn:AMomega}\end{equation}
This much larger component of the vorticity field is shown in Fig.\ref{fig:Absolute}.
The maximum magnitude of the speed of this precession  component is $\omega/2=gv_C/c^2=8 \times10^{-6}$arcsec/s, where here
$g$ is the gravitational acceleration at the altitude of the satellite.   This precession has a different signature: it  is
not cumulative, and is detectable by its variation over each single orbit, as its orbital average is zero, to first
approximation.   Fig.\ref{fig:Precession} shows   $\Delta \Theta=|\Delta {{\bf S}}(t)|/|{\bf S}(0)|$  over 
 one orbit, where, as in general,
\begin{equation}\Delta {{\bf S}}(t) =
\int_0^t dt^\prime \frac{1}{2}\vec{\omega}({\bf r}(t')) \times {\bf S}(t^\prime)
\approx \left(\int_0^t dt^\prime \frac{1}{2}\vec{\omega}({\bf r}(t'))\right) \times
{\bf S}(0).
\label{eqn:precession}\end{equation}  
Here $\Delta {{\bf S}}(t)$ is the integrated change in spin, and where
the approximation arises  because the change in
${\bf S}(t^\prime)$ on the RHS of (\ref{eqn:precession}) is negligible.   The plot in  Fig.\ref{fig:Precession}  shows
this effect to be some 30$\times$ larger than the expected GP-B errors, and so easily detectable.   This precession is
about the instantaneous direction of the vorticity $\vec{\omega}({\bf r}((t))$ at the location of the satellite, and so is
neither in the plane, as for the geodetic precession, nor perpendicular to the plane of the orbit, as for the
earth-rotation induced vorticity effect. This absolute motion induced spin precession is shown in
Fig.\ref{fig:PrecGraphic}. 

Because the yearly orbital rotation  of the earth about the sun slightly effects 
${\bf v}_C$ \cite{AMGE} predictions for four months throughout the  year are shown  in
Fig.\ref{fig:Precession}. Such yearly effects were first seen in the Miller \cite{Miller}
experiment.  

However the main new feature of this paper is the detailed prediction of the magnitude and signature  of the new
gravitational-wave induced spin precessions that GP-B is capable of detecting, but first we briefly summarise the history of
the detection of gravitational waves.

\section{ Brief History of the Detection of Gravitational Waves}

\begin{figure}[t]
\hspace{15mm}\includegraphics[scale=1.5]{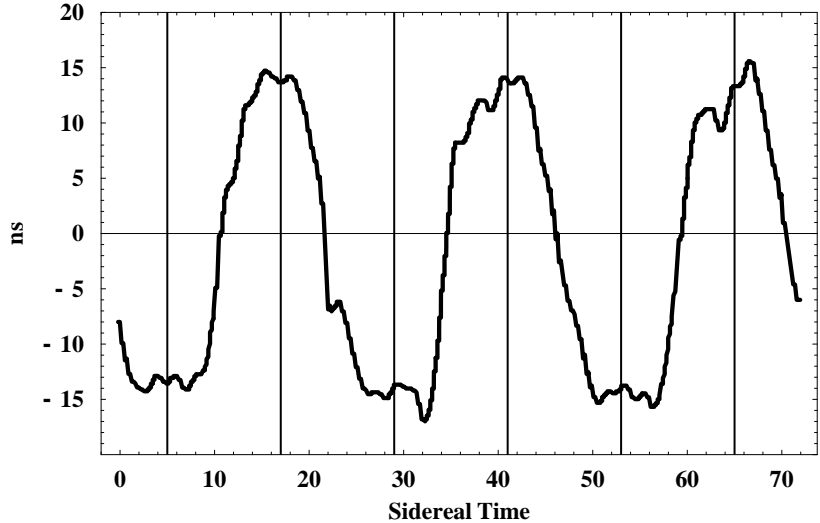}
\caption{\small{ Variations in twice the one-way travel time, in ns, for an RF signal to travel 1.5
km through a  coaxial cable between  Rue du Marais and Rue de la Paille, Brussels. 
An offset  has been used  such that the average is zero.  The definition of the sign  convention for $\Delta t$ used
by DeWitte is unclear.  The cable has a
North-South  orientation, and the data is $\pm$ difference of the travel times  for NS and SN
propagation.  The sidereal time for maximum  effect of $\sim\!\!17$hr (or   $\sim\!\!5$hr) (indicated
by vertical lines) agrees with the direction found by Miller.  Plot shows
data over 3 sidereal days  and is plotted against sidereal time in hours.  The regular  time variation is caused
by the earth rotation changing the orientation of the cable with respect to the direction of absolute linear motion.}  
\label{fig:DeWittetimes}}\end{figure}

\begin{figure}[t]
\hspace{13mm}\includegraphics[scale=1.6]{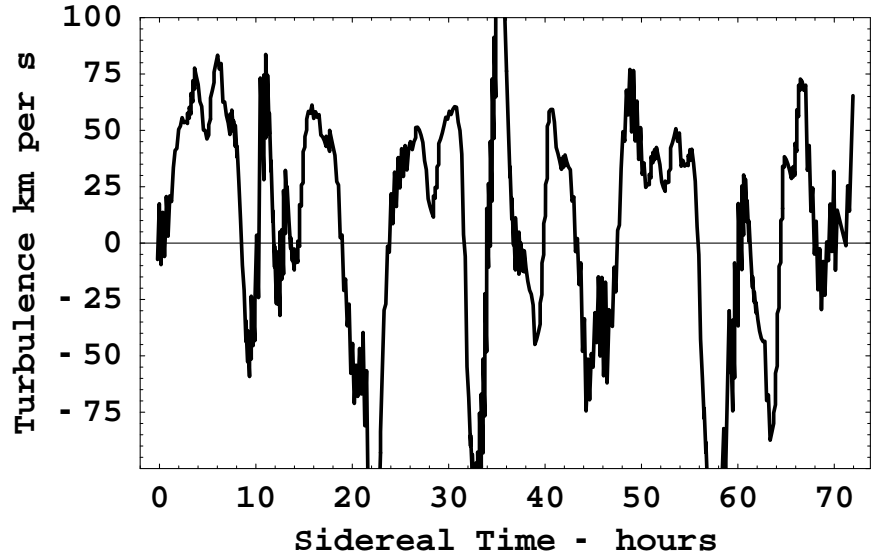}
\caption{\small{ Speed fluctuations  determined from Fig.\ref{fig:DeWittetimes} over three sidereal days, or some
70 hours. A 1ns variation in travel time corresponds approximately to a speed variation of 27km/s.  The
larger speed fluctuations  are anomalies of the fitting technique and actually arise from a fluctuation in
the cross-over time, that is, a fluctuation in the direction of the velocity.   }  
\label{fig:Turbulence}}\end{figure}

As already noted the velocity  flow-field equations
(\ref{eqn:CG4a})-(\ref{eqn:CG4b}) have wave-like solutions involving variations in both the magnitude and direction of the
velocity flow-field.  Remarkably  all the Michelson gas-mode interferometer and coaxial cable experiments showed
evidence of such wave phenomena
\cite{AMGE}. The first clear evidence was from the Miller 1925/26 experiment.  Miller offered no
explanation for these fluctuations  but in his analysis of that data he did running time averages to
remove these fluctuation effects, though some of that could also have been instrumental noise.   While
some of these fluctuations may also be partially caused by weather related temperature and pressure 
variations, the bulk of the fluctuations appear to be larger than expected from that cause alone.   Even
the original Michelson-Morley 1887 data  shows variations in  the velocity field and supports this
interpretation.    However it is significant that the non-interferometer 1991 DeWitte experiment also showed
clear evidence of turbulence  of the velocity flow field, as shown in
Fig.\ref{fig:Turbulence}.  Just as the DeWitte data agrees with the Miller data for speeds and
directions of absolute motion, the magnitude of the  fluctuations are also very similar \cite{AMGE}.  
    It therefore  becomes clear that there is
strong evidence for these fluctuations being evidence of  physical turbulence in the flow
field.  The magnitude of this turbulence appears to be  larger than that which would be
caused by the in-flow of quantum foam towards the sun, and indeed  most of this turbulence may be associated with galactic
in-flow into the Milky Way and local galactic cluster.  This in-flow turbulence is a form of 
gravitational wave and the
ability of gas-mode Michelson interferometers to detect absolute motion means that experimental
evidence of such a wave phenomena has been available for a considerable period of time.    All this
means that the new gravitational wave phenomena is very easy to detect and amounts to new physics that
can be studied in much detail.

\begin{figure}[t]
\hspace{15mm}\includegraphics[scale=1.5]{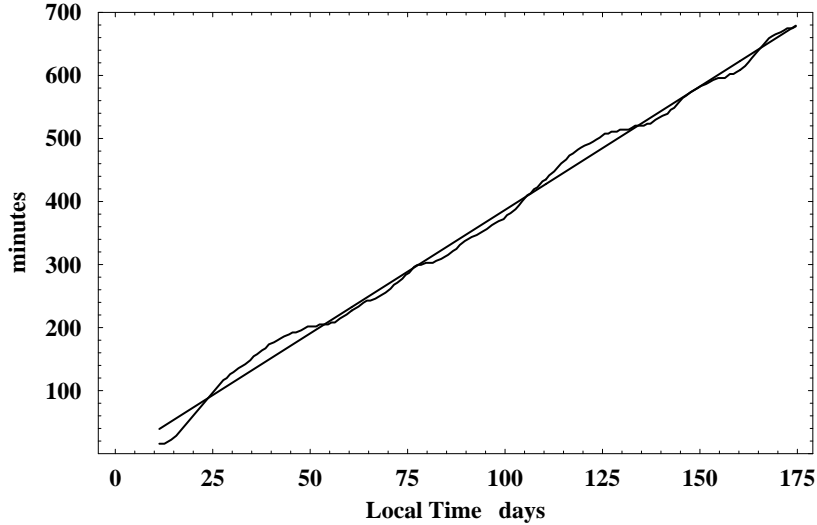}
\caption{\small{ Plot of the negative of the drift of the cross-over time between minimum and
maximum travel-time variation each day (at $\sim10^h\pm1^h$ ST) versus local solar time for some
178 days. The straight line plot is the least squares fit to the experimental data, 
 giving an average slope of 3.92 minutes/day. The time difference between a sidereal day and a solar
day is 3.93 minutes/day.    This demonstrates that the effect is related to sidereal time and not local
solar time. The fluctuations  apparent in the plot are caused by the slow changes in direction of the flow.  }  
\label{fig:DeWitteST}}\end{figure}

 The DeWitte  1991 experiment  within Belgacom, the Belgium telecommunications company, was a 
 most impressive and high quality experiment that we now understand to have detected gravitational waves  
\cite{AMGE}.   In this serendipitous discovery two sets of atomic clock in two buildings in Brussels, separated by 1.5
km, were used to time   5MHz radiofrequency signals travelling in each direction,
  through two  buried  coaxial cables, with a N-S orientation, linking the two clusters.   The atomic clocks were
cesium beam atomic clocks, and there were three in each cluster. In that way the stability of the clocks could be
established and monitored. One cluster was in a building on Rue du Marais and the second cluster was due south in a
building on Rue de la Paille.  Digital phase comparators were used to measure changes in times between clocks within
the same cluster and also in the propagation times of the RF signals. Time differences between clocks within the same
cluster showed  a linear phase drift caused by the clocks not having exactly the same frequency together with short
term and long term noise. However the long term drift was very linear and reproducible, and that drift could be
allowed for in analysing time differences in the propagation times between the clusters.

Changes in propagation times  were observed and eventually observations over  178 days were recorded. A sample
of the  data, plotted against sidereal time for just  three days, is shown in
Fig.\ref{fig:DeWittetimes}.  DeWitte recognised that the data was evidence of absolute motion but he was unaware of
the Miller experiment  and did not realise that the Right Ascension for minimum/maximum  propagation time agreed
almost exactly with Miller's direction ($\alpha=5.2^{hr}, \delta=-67^0$). In fact DeWitte expected that the
direction of absolute motion should have been in the CMB direction, but that would have given the data a totally
different sidereal time signature, namely the times for maximum/minimum would have been shifted by approximately 6 hrs. 
The declination of the velocity observed in this DeWitte experiment cannot be determined from the data as
only three days of data are available.   However assuming exactly the same declination as Miller  the speed
observed by DeWitte appears to be also  in excellent agreement with the Miller speed, which in turn is in
agreement with that from the Michelson-Morley and Illingworth experiments \cite{AMGE}.

Importantly DeWitte reported the sidereal time of the cross-over time, that is a `zero' time
in Fig.\ref{fig:DeWittetimes}, for  178 days of data.  This is plotted in Fig.\ref{fig:DeWitteST} and
demonstrates that the time variations are correlated with sidereal time and not local solar
time.  A least squares best fit of a linear relation to that data gives that the cross-over time
is retarded, on average, by 3.92 minutes per solar day. This is to be compared with the fact that a sidereal day is
3.93 minutes shorter than a solar day. So the effect is certainly  cosmological and not associated with any daily
thermal effects, which in any case would be very small as the cable is buried.  Miller had also compared his
data against sidereal time and established the same property, namely that up to  small diurnal effects 
identifiable with the  earth's orbital  motion,  features in the data tracked sidereal time and not
solar time.

If the changes in propagation time through the coaxial cable were caused solely by the 
rotation of the coaxial cable, carried along by the rotation of 
 the earth, so changing its orientation with respect to that of a uniform  absolute  velocity,
then the variations in travel time would not show the fluctuations that are evident in  Fig.\ref{fig:DeWittetimes},
 but
only the dominant regular variation over each sidereal day.   So these fluctuations, and they are much larger than any
errors that were produced by the atomic clock timing procedures, are manifestations of a genuine novel physical effect. 
Here and in
\cite {RGC,AMGE} they are interpreted as the gravitational waves predicted by the theory in Sect.\ref{section:waves}. By
fitting the data to the time variation forms expected without such fluctuations, the fluctuations themselves may
extracted and their magnitude translated into speed variations, as shown in Fig.\ref{fig:Turbulence}.

As well Torr and Kolen in  1984 also used a coaxial cable experiment to detect absolute motion \cite{AMGE}. 
Unfortunately the
cable was orientated in an E-W direction, which made it relatively insensitive to the detection of absolute motion,
as that is in a near N-S direction, though the expected travel variations were indeed observed \cite{AMGE}.  However
being in the E-W direction  the Torr-Kolen coaxial cable  is very sensitive to the directional changes associated with
the turbulence. Being almost at
$90^0$ to the direction of absolute motion, any variation in that direction produces significant effects, as indeed
reported by Torr and Kolen.   So we  have yet another experiment in which the data is such  that we have 
interpreted it as the detection of these new  gravitational waves.

\begin{figure}[t]
\hspace{5mm}\includegraphics[scale=1.7]{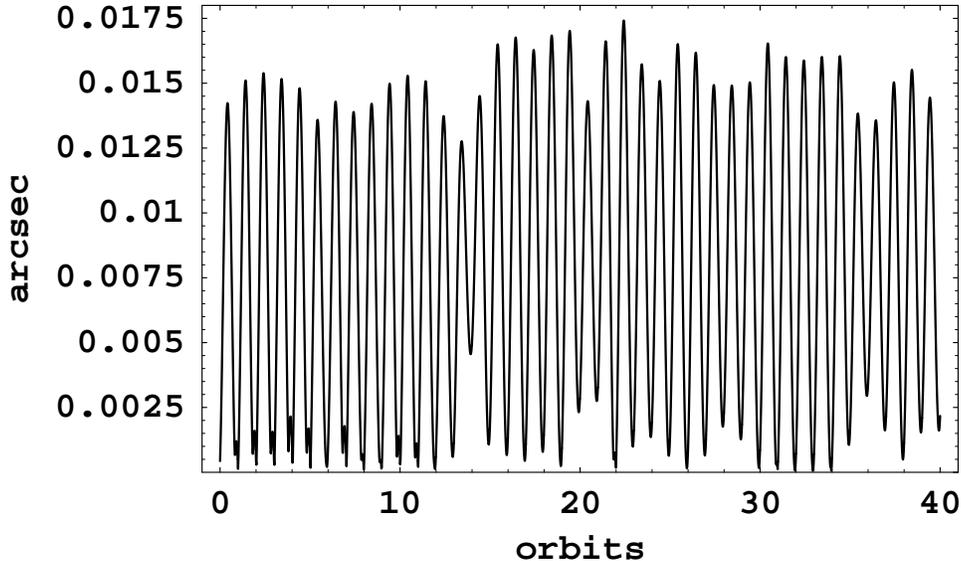}
\caption{\small Predicted  variation of the precession angle  $\Delta \Theta=|\Delta {{\bf S}}(t)|/|{\bf S}(0)|$, in
arcsec, as given by (\ref{eqn:precession}),  over 40 GP-B orbits, from the vorticity induced by the absolute linear
motion of the earth, and with the mainly galactic gravitational waves now included by means of the simulation described
in the text.  This plot shows the expected magnitude and signature of the effects of
the gravitational waves.  The  corresponding prediction for one orbit, but without the gravitational wave effect, is
shown in  Fig.\ref{fig:Precession}.  The main signature of the gravitational waves for GP-B is the  variation, from orbit
to orbit, of the maximum precession angle, as shown in this plot. The orbit time begins at location
${\bf S}$ in Fig.\ref{fig:Precession}. Predictions are for the  month of  September.    The GP-B expected angle
measurement accuracy is 0.0005 arcsec.
  \label{fig:waveprecessions}}\end{figure}

\section{ Detection of Gravitational Waves  by GP-B}

The novel gravitational waves displayed in Fig.\ref{fig:Turbulence} are significant in magnitude compared to the
average absolute motion speed of some 430 km/s. For that reason they  have a significant effect upon the vorticity
field, and consequently  of the GP-B spin precessions. To give a first indication of the expected size of these
wave induced spin precessions we have  used the speed magnitude fluctuations in Fig.\ref{fig:Turbulence} by
simply adding these fluctuations to the speed of 430 km/s in the direction RA $ = 5.2^{h}$,
Dec$ = -67^0$, but without allowing for any fluctuation in direction.  Then the spin precession in
(\ref{eqn:precession})  was integrated over 40 orbits of the satellite, with the results shown in
Fig.\ref{fig:waveprecessions} for September.  This figure should be compared with the form in Fig.\ref{fig:Precession}
  for only one orbit, and without the gravitational wave effect.  We see the magnitude and signature of the gravitational
wave induced spin precessions, including the effect that the absolute motion induced precessions now no longer return to
zero after each orbit.   Whether this effect persists over many months  is not possible to predict, but if it did persist
then it would seriously interfere with, in particular, the observation of the earth-rotation  induced spin precession,
which is cumulative but very small. 

There is a technical difficulty in observing the effects shown in Figs.\ref{fig:Precession} and
\ref{fig:waveprecessions}, namely that the guide star is not visible to the on-board GP-B telescope   
during the `mid-part' of each orbit, as is apparent from Fig.\ref{fig:Absolute}.

\section{ Conclusions}

By using the gravitational wave effects revealed by the DeWitte coaxial cable absolute linear motion experiment we
have been able to simulate the effects of these waves upon the spin precession of the GP-B gyroscopes, in order
that the signature of these effects may be recognised in the GP-B data.  Both the absolute linear motion and the 
new gravitational waves are properties of the new theory of space and gravity. These effects are absent from
General Relativity, and  which, most significantly, has had only one observational test, namely the binary pulsar
slow-down, so far, of its dynamical properties, as distinct from the various tests of the geodesic equation.  The
new theory of gravity has explained a number of observational and laboratory effects, but in particular it has
explained the so-called `dark matter' effect as a dynamical property of space itself; so it is not caused by any
form of `matter'. So the `dark matter' effect was a fatal flaw of  General Relativity. As well the spacetime
ontology, as distinct from the mathematical formalism, is also seriously flawed by the fact that so far at least
seven experiments have detected absolute linear motion. The GP-B experiment will also detect evidence of this
absolute motion because of the vorticity associated with that motion, that is, caused by the earth passing through
space, which includes also a much smaller vorticity caused by the earth's rotation. This component is identical to
the vorticity caused by the earth's rotation within  General Relativity.  So General Relativity is in the
paradoxical situation of permitting absolute rotational motion but banning absolute linear motion. One major
outcome of the GP-B experiment will be the  definitive resolution of this longstanding paradox.

\end{document}